\documentstyle[seceq,letter,epsf]{ptptex}

\title{
A Model for the
$^3$He($\vec d$, p)$^4$He Reaction at Intermediate Energies\footnote{
Dedicated to Prof. Shinsho Oryu on the occasion of his  60th birthday}
}

\author{
H. Kamada
\footnote{present address:
Institut f\"ur Strahlen- und Kernphysik der Universit\"at  Bonn
Nussallee 14-16, D-53115 Bonn, Germany, E-mail address: kamada@tp2.ruhr-uni-bochum.de 
 }$^{,***),****)}$,
W. Gl\"ockle
$^{***)}$,
H.  Wita\l a
$^{*****)}$,
S. Gojuki
$^{****)}$
}

\inst{
${}^{***)}$
Institut f\"ur  Theoretische Physik II,
         Ruhr-Universit\"at Bochum, D-44780 Bochum, Germany
\\
${}^{****)}$
Department of Physics, Faculty of Science and Technology,
Science University of Tokyo, Noda, 278-8510 Chiba, Japan 
\\
${}^{*****)}$
Institute of Physics, Jagellonian University,
PL-30059 Cracow, Poland
}

\recdate{
\today
}

\abst{
Polarization correlation coefficients have been  measured  at  RIKEN for the
$\vec {^3He}(\vec d,p)^4He$  reaction at intermediate energies.
We propose a model for the  ($\vec d$, p) reaction mechanism  using the $pd$ elastic
scattering amplitude
which is rigorously determined  by  a Faddeev calculation and using modern NN forces.
Our theoretical predictions for deuteron polarization observables
$A_y$, $  A_{yy}$, $A_{xx}$  and  $A_{xz}$ at $E_d$=140, 200 and 270 MeV agree qualitatively in shape with the
experimental data for the reaction $ {^3He}(\vec d,p)^4He$.
}

\begin{document}

\maketitle

\section{Introduction}

The measurement\cite{UESAKA} of  the  $\vec {^3He}$ ($\vec d$, p)$^4He$ reaction 
at RIKEN 
aimed at an investigation of 
the high momentum components of the deuteron wave function and the d-state admixture linked to them.
High precision data resulted for the polarization observables $A_y$, $A_{yy},A_{xx},C_{y,y}$ and $C_{x,x}$.
Out of them the linear combination  $C_\parallel= 1+{1 \over 4} (A_{yy}+A_{xx})+{3 \over 4} (C_{y,y}+C_{x,x}) $ 
has been formed\cite{UESAKA}.
The Dubna and Saturne groups also obtained the polarization correlation coefficient  $C_\parallel$  built in this case 
from  the
measurements  of $T_{20}$ and $\kappa_0$ in d + p backward scattering\cite{DPBS} 
and from the inclusive deuteron  breakup process\cite{DIB}.
The polarization correlation coefficient $C_\parallel$ at forward angles of the outgoing proton  
is directly related to  the ratio of deuteron wavefunction components
if one uses the plane wave impulse approximation (PWIA) :
\begin{eqnarray}
C_\parallel (PWIA) \equiv  { 9 \over 4 } {  w ^2 (k_{pn} ) \over { u ^ 2 ( k _{pn} ) + w^2 ( k_{pn}) }}
\label{EQ1}
\end{eqnarray}
Here $u$ and $w$ are the S-, D-wave components of the deuteron wavefunction, and $k_{pn}$ a kinematically fixed 
relative momentum of the $pn$ pair.
These PWIA calculations are very poor in relation to the data\cite{UESAKA}.
This is shown in table \ref{table1} for $C_\parallel$. There we also exhibit the different D-state probabilities 
for the modern realistic NN potentials, CD-Bonn\cite{BONN}, AV18\cite{Argonne} and Nijmegen I,II and 93\cite{NIJM}.
Clearly one needs a better calculation for the analysis of the $^3\vec He$($\vec d $,p)$^4$He reaction. 

A theoretical analysis has been reported by the SUT group \cite{ORYU} based on a  $^3$He-n-p and d-d-p 
three-cluster model.
However, the evaluations performed up to now in this model lead only to a tiny deviation from the PWIA calculations
 just mentioned. 
Recently, the Hosei group \cite{HOSEI} analyzed $T_{20}$ and $\kappa_0 $ with the  $^3$He-n-p cluster model by 
an analogy between $^3$He and the proton (T=1/2, S=1/2). They conclude that PWIA describes the 
global features of the experimental 
data. 

In this  letter we would like to 
introduce again a 3N model, which when evaluated correctly leads to a great similarity of various polarization 
observables to the ones found in the reaction $^3\vec He$($\vec d $,p)$^4$He.

\begin{table}
\caption{\label{table1}
The polarization correlation coefficient $C_\parallel$ according to Eq. (\ref{EQ1})  
in a simple PWIA model for different NN potentials and from an experimental 
value for the reaction $^3\vec He$($\vec d$,p)$^4$He. 
We also show the corresponding deuteron  D-state probabilities.
}
\begin{tabular} {c@{\hspace{5mm}}|c@{\hspace{5mm}}|c@{\hspace{5mm}}}
\hline \hline
Potential &   $C_\parallel (PWIA) $ &  D-state Probability (\%) \\
\hline
CD-Bonn \cite{BONN}&  0.645    &    4.86                \\
AV18 \cite{Argonne} &  0.722      &   5.78             \\
Nijmegen 93 \cite{NIJM} &   0.710   &   5.76               \\
Nijmegen I\cite{NIJM} &    0.712      &  5.68                 \\
Nijmegen II\cite{NIJM} &    0.726     &  5.65       \\
\hline
exp. \cite{UESAKA} & 0.223$\pm$ 0.044(statistical)$\pm$ 0.037(systematic) & - \\
\hline
\end{tabular}
\end{table}

\section{Model}

\begin{figure}[hbtp]
\begin{center}
\mbox{\epsfysize=20mm\epsffile{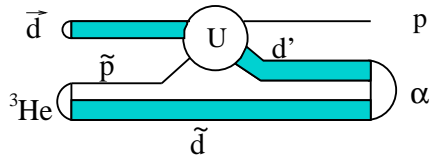}}
\bigskip
\caption{ Diagram of the reaction mechanism 
}
\label{FIG1}
\end{center}
\end{figure}

For the $^3$He($\vec d$,p)$^4$He reaction we assume a model which is based on a three-body reaction process.
This is shown in Fig. \ref{FIG1}.
The wavefunctions for $^3$He and $^4$He take on maximal values if the momenta of the subclusters are zero 
in their respective rest systems.
These are for $^3$He the momenta of p and d and for $^4$He the momenta of the two deuterons.
This means that for the moving nuclei the subcluster momenta should be equal.
Therefore to form the $\alpha$-particle with highest probability in the picture of Fig. \ref{FIG1} 
one has to assume that the two deuterons, $d'$ and $\tilde d$,
 have equal momenta. Likewise for $^3$He one has to assume
that the proton and deuteron, $\tilde p$ and $\tilde d$, 
 have equal momenta. This turns out to be kinematically inconsistent. 
Therefore we  make a choice and assume that only  the two deuterons forming the $\alpha$ particle have
equal momenta. We justify this choice by the larger binding energy of the $\alpha$ particle.

\begin{figure}[hbtp]
\begin{center}
\mbox{\epsfysize=25mm\epsffile{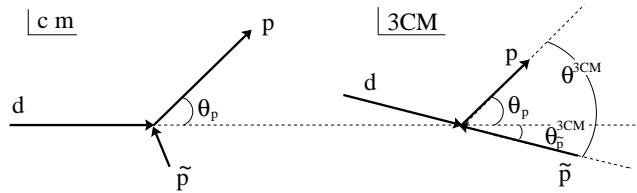}}
\bigskip
\caption{ Scattering angles for the  5-body (cm) and 3-body (3CM) center of mass systems.
}
\label{FIG2}
\end{center}
\end{figure}

It is easy to see that our basic assumption 
\begin{eqnarray}
\vec k _{\tilde d} = \vec k _{d '} 
\end{eqnarray}
fixes the kinematics uniquely. It follows by simple kinematical arguments that 
\begin{eqnarray}
\vec k_{\tilde p } ^{lab} = { 1 \over 2} \vec k _p ^{cm} - {2 \over 5} \vec k _d ^{lab} = -\vec
k_{\tilde d} ^{lab}
\end{eqnarray}
Here the superscripts $lab$ and $cm$ denote the laboratory and 5-body cm systems, respectively.
Further the total momentum of the picked up proton and the incoming deuteron in the lab system is
\begin{eqnarray}
\vec K = { 1\over 2 } \vec k_p ^{cm} + { 3 \over 5} \vec k_d ^{lab}
\end{eqnarray}
Also we get the momentum of the picked up proton in the 3-body center of mass system (3CM) as
\begin{eqnarray}
\vec k _{\tilde p} ^ {3CM} = { 1 \over 3 } \vec k _{p}^ {cm} - { 3 \over 5} \vec k_d ^{lab} 
\end{eqnarray}
and the 3CM energy as 
\begin{eqnarray}
E_{3CM} = { 3 \over 4 m} ( \vec k _{\tilde p} ^{3CM} ) ^2
\label{E3CM}
\end{eqnarray}
We show in Fig. \ref{FIG2} the relevant kinematics for the cm and the 3CM systems. 
From the relation 
\begin{eqnarray}
\vec k _{\tilde p} ^{3CM}  = { 2 \over 5}  \vec k _{p} ^{3CM} - { 3 \over 5} \vec k_d ^{lab}
\end{eqnarray}
it follows under our condition, that the angles shown in Fig. \ref{FIG2} are related as 
\begin{eqnarray}
\theta ^{3CM} = \theta _p ^{3CM} - \theta _{\tilde p } ^{3CM} 
\label{THETA3CM}
\end{eqnarray}
(note that $\theta_p \equiv \theta _p ^{3CM} = \theta_p ^{cm}$ ).
The dependence of $E_{3CM}$ on $\theta_p ^{cm}$ is illustrated in Fig. \ref{FIG3} for 3 deuteron 
energies. The scattering angle $\theta^{3CM}$ is shown against $\theta_p ^{cm}$ in Fig. \ref{FIG4}
again for the same 3 deuteron energies.

Our claim is now that  
$ 
{\cal O} (E_d, \theta_p ^ {cm} )  \approx  {\cal O}_ {pd}( E _{3CM},  \theta   ^{3CM}) 
$ 
where ${\cal O}_{pd}$ are the elastic pd deuteron polarization observables and $\cal O$ the ones for the reaction 
$^3\vec He$($\vec d $, p)$^4$He. 

Before calculating these 3N  observables we introduce one   more approximation.
Looking at Fig.\ref{FIG3} we see that $E_{3CM}$ varies with $\theta_p ^{cm}$ and consequently for each 
$\theta_p^{cm}$ one would have to solve the 3N Faddeev equation. We  avoided that for that qualitative 
investigation and have chosen available Faddeev results at three energies which lie in the three energy bands for 
$ 0 < \theta_p ^{cm} < 40 ^\circ $. They are $E_{3CM}$= 66.7, 100, 133 MeV corresponding to $E_d$= 140, 200,
 270 MeV, respectively.

        \begin{figure}[htb]
            \parbox{\halftext}{
\input{Ene_tex}
 \caption{\label{FIG3}Effective $E_{3CM}$  energies as a function of the proton scattering angle $\theta_p ^{cm}$.
The solid, dashed and short dashed lines are for  $E_d$=140, 200 and 270 MeV, respectively.   }
}
            \hspace{8mm}
            \parbox{\halftext}{
\input{Angle_tex}
                \caption{\label{FIG4} Effective scattering angle $\theta ^{3CM} $ as a function of the proton scattering angle
 $\theta_p ^{cm}$ for the deuteron energies as in  Fig. \ref{FIG3}. }}
        \end{figure}

\section{Results}

As NN potential we used AV18 in the Faddeev calculations. 
The operator $U$ for elastic scattering has the 
form (see, for instance, \cite{GLOECKLE})
 \begin{eqnarray}
U =  P G_0^{-1}  + P T 
\end{eqnarray}
where $G_0$, $P$ and $T$ are the free 3N propagator, permutation operators and a partial 3N break-up operator,
which is determined by a Faddeev equation. The first term, the famous nucleon exchange term, is essentially related
to the PWIA mentioned in introduction. 
In order to see the importance of solving the Faddeev equation correctly and not just replacing $U$ by 
$P G_0 ^{-1}$  we compare the corresponding predictions for $A_{yy}$, $A_{xx}$ and $A_{xz}$ in Figs. 
 \ref{FIGAYY2}-\ref{FIGAXZ2}. 
We see large differences especially above about 15 degrees. Trivially $A_y$ is identically zero using only 
the real term $P G_o ^{-1}$.

        \begin{figure}[htb]
            \parbox{\halftext}{
\input{Ayy2_tex}
 \caption{\label{FIGAYY2} Tensor analyzing power $A_{yy}$ in elastic pd scattering at $E_{CM}$=133 MeV. 
The solid (dashed) line is calculated from $U$ ($P G_0 ^{-1}$).
 The data point for $^3\vec {He}$($\vec d$,p)$^4$He reaction
 is from \protect \cite{UESAKA}.  }}
            \hspace{8mm}
            \parbox{\halftext}{
\input{Axx2_tex}
                \caption{ \label{FIGAXX2}  The same as Fig. \ref{FIGAYY2} for $A_{xx}$. }}
        \end{figure}

\begin{figure}[hbtp]
\begin{center}
\input{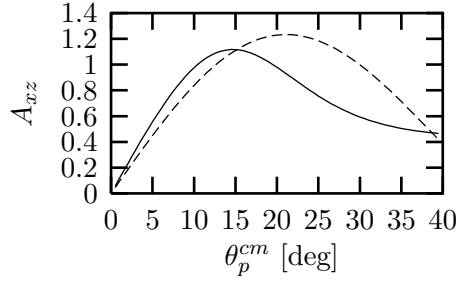}
\bigskip 
\caption{\label{FIGAXZ2} The same as Fig. \ref{FIGAYY2} for $A_{xz}$.  There is no data point. 
}
\end{center}
\end{figure}

        \begin{figure}[htb]
            \parbox{\halftext}{
\input{Ay_tex}
 \caption{ \label{FIGAY}    The deuteron vector analyzing power $A_{y}$ for the pd elastic scattering for $E_{3CM}$=66.7(solid),
 100(long dashed)
 and 133(short dashed) MeV as a function of $\theta_p ^{cm} $, corresponding to the 
 $^3$He($\vec d$,p)$^4$He reaction for $E_d$=140, 200, 270
MeV, respectively.    }}
            \hspace{8mm}
            \parbox{\halftext}{
\input{Ayy_tex}
                \caption{ \label{FIGAYY}   The same as in Fig. \ref{FIGAY} for $A_{yy}$.   The data point for $^3\vec {He}$($\vec d$,p)$^4$He
 reaction is from \protect \cite{UESAKA}.  }}
        \end{figure}

        \begin{figure}[htb]
            \parbox{\halftext}{
\input{Axx_tex}
 \caption{\label{FIGAXX}   The same as in Fig. \ref{FIGAY} for $A_{xx}$.   The data point for $^3\vec {He}$($\vec d$,p)$^4$He
 reaction is from \protect \cite{UESAKA}.  }}
            \hspace{8mm}
            \parbox{\halftext}{
\input{Axz_tex}
                \caption{ \label{FIGAXZ} The same as in Fig. \ref{FIGAY} for $A_{xz}$. }}
        \end{figure}

The predictions of the full Faddeev solution are shown in Figs.  \ref{FIGAY}-\ref{FIGAXZ} at
 $E_{3CM}$=66.7, 100 and 133 MeV, respectively.
This should be compared to recent data\cite{TAIWAN}.  We see a behavior qualitatively similar to those data, 
especially for $A_y$. For the $A_y$ data the minima shift to smaller $\theta_p ^{cm} $ value with increasing 
energy like in Fig. \ref{FIGAY}. Also for $A_{yy}$ the qualitative behavior is similar in our model and the 
data, especially at the highest energy. For $A_{xx}$ the shapes are again very similar. 
In Fig. \ref{FIGAYY} and \ref{FIGAXX} we include one data point from \cite{UESAKA}. This shows that 
our absolute values are too high. For $A_{xz}$ shown in Fig. \ref{FIGAXZ} there are not yet data.

\section{Summary and Outlook}

We assumed that the reaction $^3\vec He$($\vec d$,p)$^4$He at forward angles is mainly driven by elastic 
pd scattering. In this model the deuteron picks up a proton from $^3$He, scatters elastically and combines 
then again with the spectator nucleons to an $\alpha$ particle. 
Our main assumption is that the momentum of the scattered deuteron equals the spectator momentum of the
deuteron in $^3$He. This leads to a high probability to form the final $\alpha$-particle.
The resulting spin-observables are in 
astonishingly good 
qualitative agreement with the data. Important thereby is, that the elastic pd amplitude 
is a full solution of the 3N Faddeev equation and not only a simple PWIA expression. This model should be 
generalized by the mechanism that also a neutron from $^3$He can be picked up. In this case one has to use 
the nd break-up amplitude. Since the polarization of $^3$He is carried by more than 90 \% by the neutron 
this second mechanism is of course mandatory for a description of $C_{x,x}$ and $C_{y,y}$. The proton pick-up
alone is too poor for those spin correlation observables. 
Also we neglected the momentum distributions of the proton in $^3$He 
and of the deuteron in the $\alpha$ particle. As an additional improvement the spin of the deuteron 
should be properly rotated for the deuteron polarization observables.

Based on the promising qualitative results achieved it appears worthwhile to improve and enrich the model along 
the lines mentioned.  

\section*{Acknowledgements}

This letter is dedicated to Prof. Shinsho Oryu on  the occasion of his  60th's birthday.
Authors would like to thank Prof. Hideyuki Sakai, Dr. Tomohiro Uesaka, Mr. Yositeru Satou 
and Ms. Kimiko Sekiguchi for 
fruitful discussions in RIKEN.
This work was supported by the Deutsche Forschungsgemeinschaft. 
The numerical calculations have been performed on the CRAY T90 of the 
John von Neumann Institute for Computing 
in J\"ulich, Germany.

\end{document}